\documentclass{cernyrep} 
\usepackage{texnames}
\usepackage[T1]{fontenc}
\usepackage[bookmarks, colorlinks=true, linktoc=page, pdftex, linkcolor=blue, citecolor=blue, urlcolor=blue]{hyperref}
\pdfoutput=1

\usepackage{subfigure}

\usepackage{textcomp} 
\usepackage{amsmath,amssymb}
\usepackage{textgreek}
\usepackage{varwidth}
\usepackage{xcolor}

\begin{document}

\title{The CLIC Detector Concept}
\author{Florian Pitters \\On behalf of the CLICdp collaboration}
\institute{CERN, Geneva, Switzerland \\Vienna University of Technology, Austria}

\maketitle


\begin{abstract}
The Compact Linear Collider (CLIC) is a concept for a future linear collider that would provide e$^+$e$^-$ collisions at up to 3\,TeV. The physics aims require a detector system with excellent jet energy and track momentum resolution, highly efficient flavour tagging and lepton identification capabilities, full geo\-metrical coverage extending to low polar angles, and timing information of the order of nanoseconds to reject beam-induced background. To deal with these requirements, an extensive R\&D programme is in place to overcome current technological limits. The CLIC detector concept includes a low-mass all-silicon vertex and tracking detector system and fine-grained calorimeters designed for particle flow analysis techniques, surrounded by a 4\,T solenoid magnet. An overview of the requirements and design optimisations for the CLIC detector concept is presented.
\end{abstract}



\section{Introduction}
The LHC has pushed the energy frontier to new heights. For the precision frontier to maintain the pace, a high-energy lepton collider is needed. The Compact Linear Collider, CLIC, is a proposed concept for such a lepton collider. The study, hosted by CERN, aims to provide e$^+$e$^-$ collisions at up to $\sqrt{s} = 3$\,TeV in the post-LHC era.

CLIC offers a unique sensitivity to particles produced in electroweak interactions. The rich physics programme includes precision measurement of Higgs and top quark properties as well as direct and indirect searches of physics beyond the Standard Model. An overview of the physics potential of CLIC is given in Ref. \cite{clic2012-03} and a more detailed view on Higgs physics can be found in Ref. \cite{clic2016-01}.


\section{The CLIC accelerator}
For a linear collider to be able to accelerate particles to multi-TeV energies at reasonable lengths, the accelerating structures must operate at very high electrical field gradients. This excludes the use of super\-conducting RF structures because their maximum gradient is intrinsically limited by the critical field of the used material. Normal conducting cavities, however, have been shown to hold accelerating gradients of 120\,MV/m with reasonable breakdown rates when operated at a frequency of several GHz \cite{clic2012-07}. For CLIC, the goal is to obtain gradients of 100\,MV/m at a frequency of 12\,GHz.

\begin{figure}
      \centering
      \includegraphics[width=0.95\textwidth]{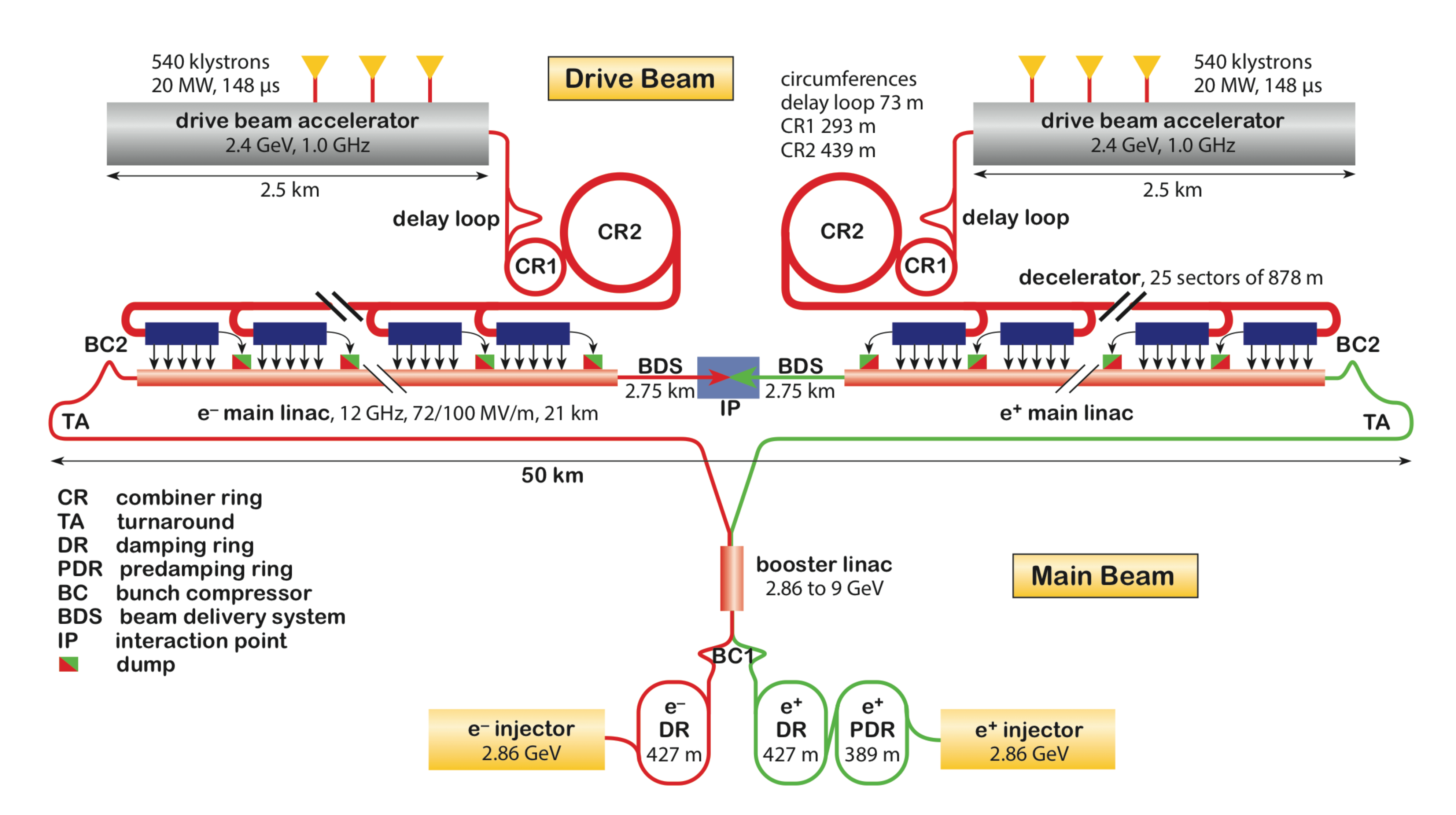}
      \caption{Two-beam acceleration scheme for CLIC, used to accelerate electrons and positrons to 3\,TeV. A low-energy but high-intensity drive beam is bunched and decelerated to power the cavities for the main beam at the desired frequency of 12\,GHz. This allows for accelerating gradients of 100\,MV/m. Figure is taken from Ref. \cite{clic2012-07}.}
      \label{fig:two_beam}
\end{figure}

To power the accelerating cavities efficiently at this frequency, CLIC is based on a novel two-beam acceleration scheme, shown in Fig. \ref{fig:two_beam}. The idea is to use a high-intensity but rather low-energy electron beam, the so-called drive beam, and restructure it into 12\,GHz bunches via a series of delay loops and combiner rings. This beam is then decelerated in dedicated cavities and the extracted 12\,GHz power is transferred via wave guides to the accelerating cavities of the main electron and positron beams. The resulting beam structure shows a bunch spacing of 0.5\,ns with 312 bunches making up one train. The repetition rate is 50\,Hz.

To achieve the desired luminosity of $5.9 \times 10^{34}$\,cm$^{-2}$s$^{-1}$, the beam sizes at the interaction point are focused to $\sigma_x \approx 40$\,nm and $\sigma_y \approx 1$\,nm at 3\,TeV. The strong focusing, together with the beam structure of the two-beam acceleration scheme, creates very high charge densities. As a result, beamstrahlung reduces the energy of individual electrons and positrons. Collisions therefore take place over a wide range of energies. At 3\,TeV, 35\% of all collisions are within 1\% of the nominal $\sqrt{s}$ value \cite{clic2012-03}. This distribution, called the luminosity spectrum, must be measured and deconvoluted in every physics study.

Due to the wide range of the physics programme, it is convenient to build CLIC in several energy stages, each one optimised for a certain part of the programme. The current baseline foresees three stages, of 380\,GeV, 1500\,GeV, and 3000\,GeV. These energy stages have recently been revisited and updated \cite{clic2016-04}.


\section{Detector requirements}
The design requirements for the detector are driven by the desired precision of the physics measurements. The track momentum resolution determines the feasibility and precision of many physics studies, \eg via the reconstruction of the di-muon invariant mass. The aim is set to $\sigma_{p_T}/p_T^2 \approx 2 \times 10^{-5}$\,GeV$^{-1}$. Another crucial parameter is the jet energy resolution. A value of $\sigma_\text{E}/E \approx 3.5$\% for jet energies above 100\,GeV allows for about $2.5\sigma$ separation of W and Z candidates in hadronic decays \cite{clic2012-03}. Moreover, efficient identification of secondary vertices for flavour tagging of heavy quark states is needed. The derived requirement is a transverse impact parameter resolution of $\sigma_{r\phi} \approx (5 \oplus 15/p[\!\UGeV]\sin^\frac{3}{2}\theta)$\,\textmu m.

In addition to the physics-driven detector requirements, the strongly focused beams and short bunch spacing set requirements on pile-up mitigation and background suppression. Incoherent e$^+$e$^-$ pairs and low $p_T$ hadronic jets from the beamstrahlung interactions are the dominant background processes. Owing to the boosted nature of this production, tagging of very forward particles plays an important role in improving the efficiency of background identification. Therefore, a large geometrical coverage in the forward region is desired. Time stamping capabilities of 1--10\,ns and a high granularity throughout the detector are also required.


\section{The CLIC detector}
\begin{figure}
      \centering
      \includegraphics[width=0.95\textwidth]{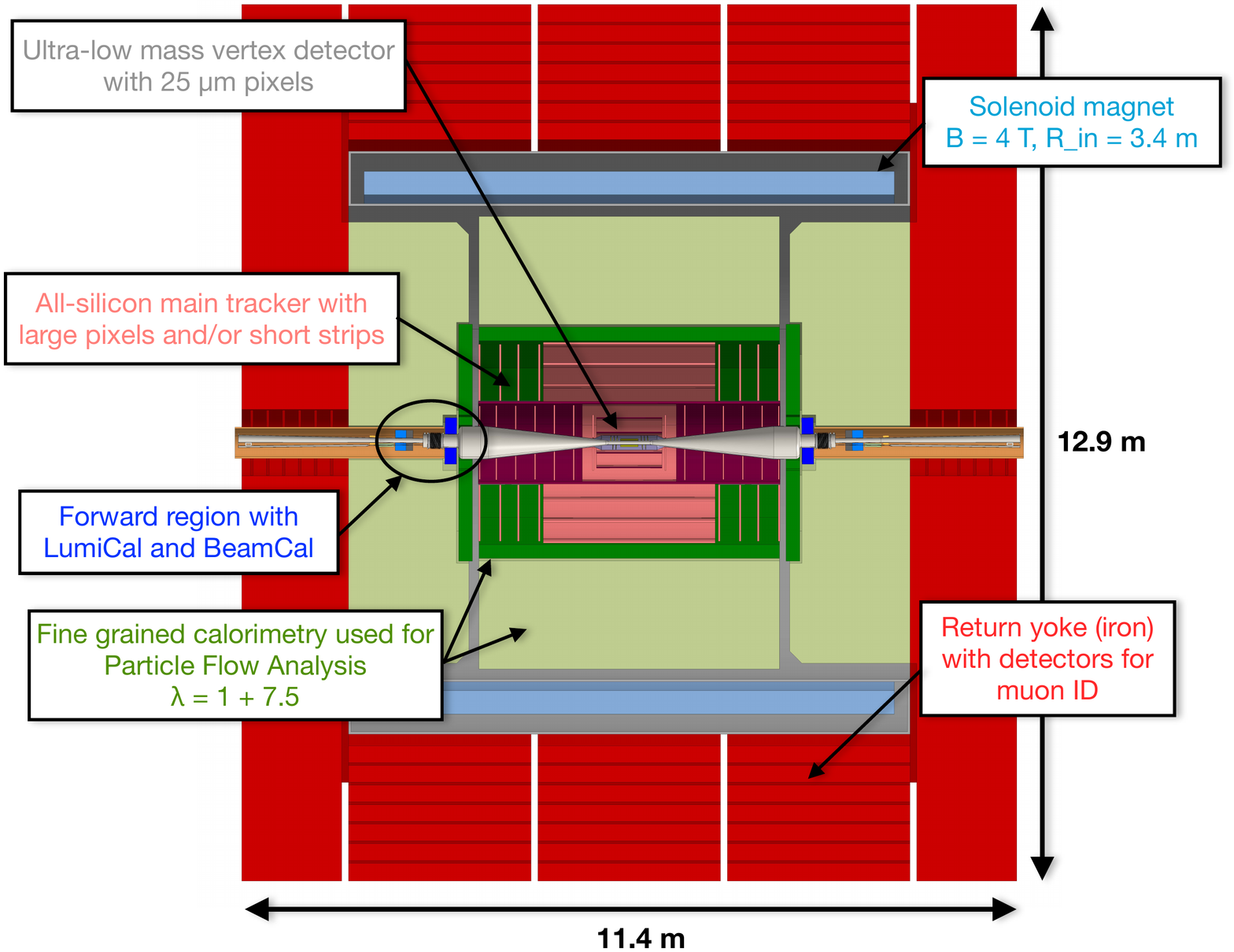}
      \caption{Top view of CLIC detector model}
      \label{fig:detector}
\end{figure}

Starting from the ILD and SiD concepts \cite{ild2010,sid2009}, adaptations were made towards CLIC's specific requirements. The resulting two concepts were consequently merged into one optimised detector model (Fig.~\ref{fig:detector}). The innermost layer of the CLIC detector is a low-mass vertex detector followed by an all-silicon tracker. The detector is optimised for particle flow algorithms; therefore, the calorimeters must be placed inside the magnet and as close to the tracker as possible. The electromagnetic calorimeter has a depth of $23X_0$ and the hadronic calorimeter is $7.5\lambda_I$ deep. A superconducting solenoid that produces a 4\,T magnetic field and an iron return yoke with interleaved muon chambers are located on the outside. The forward region is equipped with two additional calorimeters for extended geometrical coverage and luminosity measurements. The overall dimensions of the detector are 11.4\,m in length and 12.9\,m in height.

\subsection{Vertex detector}
\begin{figure}
      \centering
      \includegraphics[width=0.70\textwidth]{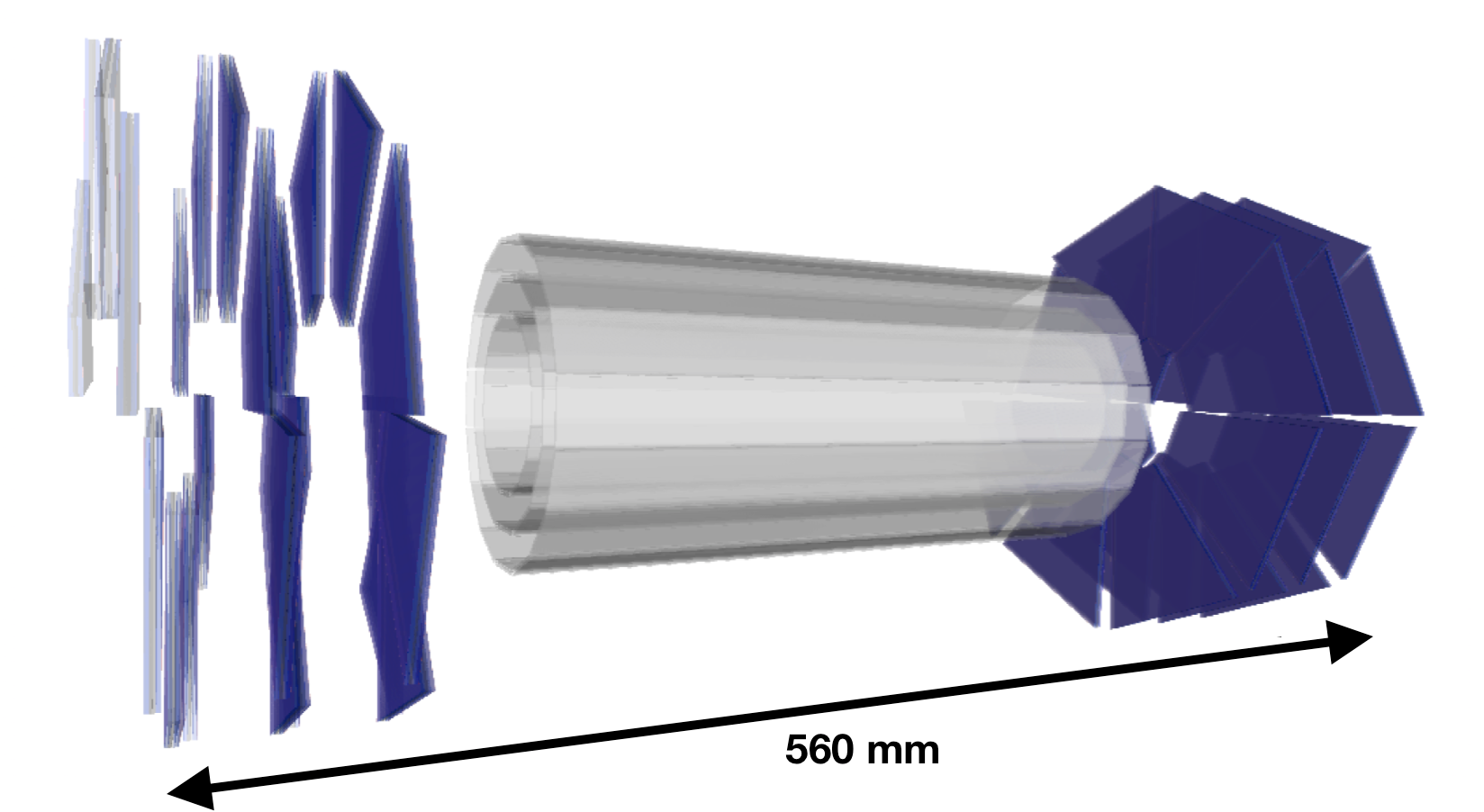}
      \caption{Geometry of CLIC vertex detector. A barrel design with spiral endcaps, together with power pulsing of the electronics, allows for forced gas flow cooling. This is necessary to fulfil the strict material budget.}
      \label{fig:vertex}
\end{figure}

The requirements for the vertex detector are determined mainly by the desired momentum resolution and flavour tagging capabilities as well as the need for efficient background suppression. To achieve the aims outlined in the previous section, the goal is to reach a single-point resolution of $\sigma_{x,y} \approx 3$\,\textmu m and time stamping capabilities of better than 10\,ns. In addition, for occupancy reasons, the pixel pitch should not exceed $25 \times 25$\,\textmu m$^2$. Another challenge comes from the low material budget. The goal of 0.2\%$X_0$ per layer translates to an equivalent of roughly 200\,\textmu m of silicon for sensor, readout, cooling, support, and cabling. The geometry of the vertex detector is shown in Fig. \ref{fig:vertex}. It is designed in three double layers and has an overall length of 560\,mm. The innermost barrel layer is located 31\,mm from the interaction point.

A concept based on hybrid silicon pixel detectors is under development to fulfil the requirements. Due to the low material budget, the sensors and readout ASICs are both foreseen to be thinned to 50\,\textmu m thickness. As sensors, either capacitively coupled HV-CMOS sensors or bump-bonded active-edge planar sensors are considered. The capacitive coupling of the HV-CMOS sensors to the readout ASIC is realised via a layer of glue. CLICpix, a demonstrator chip in 65\,nm technology with $25 \times 25$\,\textmu m$^2$ pixel pitch, has been developed for the readout. The chip enables simultaneous time and energy measurements via time-over-threshold and time-of-arrival counters. An improved version, CLICpix2, is currently in the final verification stage.

To avoid the need for liquid cooling, the layout of the vertex detector is optimised for low power dissipation. This is achieved via power pulsing of the electronics: taking advantage of the pulsed beam structure, most of the electronics is powered down after every particle train and only switched back on a few \textmu s before the next. Together with an optimised spiral geometry in the endcaps, this allows for forced air flow cooling.

\subsection{Tracking detector}
For the tracker, a single-point resolution of $\sigma_{r\varphi} \approx 7$\,\textmu m in the $r\varphi$ plane and time stamping capabilities of better than 10\,ns are needed to achieve the desired momentum resolution and background suppression. Moreover, track reconstruction requires that occupancies should be kept below 3\%. An all-silicon tracker is under development to meet these requirements. The maximum cell pitches needed are between 1\,mm and 10\,mm in the $z$ direction, depending on the position inside the detector. In the $r\varphi$ plane, a 50\,\textmu m pitch is  needed to reach the required single-point resolution. Several monolithic silicon pixel technologies are being evaluated as possible sensor candidates.

The mechanical design is divided into inner and outer regions with separate supports. The inner tracker consists of three barrel layers and seven forward discs; the outer tracker of three barrel layers and four forward discs. A radius of 1.5\,m together with a 4\,T magnetic field were chosen to achieve the required momentum and jet energy resolution at a reasonable field strength and bore radius. The variations of the magnetic field across the tracker volume are below 9\%. The overall length of the tracker is 4.4\,m.

A lightweight support is needed to meet a total material budget of ${\approx}1.5\% X_0$ per layer. To achieve this goal, a carbon fibre support frame is envisaged. A prototype to validate this concept has been built and is under evaluation.

\subsection{Calorimetry}
To achieve a jet energy resolution of approximately 3.5\%, the CLIC calorimeter is optimised for particle flow calorimetry. The basic principle here is to improve the jet energy resolution by resolving energy depositions of the individual particles in a jet and using the most precise energy measurement available for those particle types. For example, the energy measurement for charged hadrons is typically far more precise in the tracker than in the hadronic calorimeter. The jet energy resolution is then strongly depend\-ent on error contributions coming from wrongly associated depositions \cite{marshall2013}. These contributions are represented by a newly introduced confusion term that contributes to the uncertainty of the jet energy resolution. Any detector dedicated to particle flow calorimetry must be carefully optimised to minimise this term. Effectively, this approach changes the problem of summing up energy depositions in the calorimeters into a problem of pattern recognition. Dedicated software to reconstruct the particle flow is necessary for the analysis. For CLIC, Pandora PFA is used \cite{thomson2009,marshall2013-2}.

\subsubsection{Electromagnetic and hadronic calorimeter}
\begin{figure}
      \centering
      \includegraphics[width=0.70\textwidth]{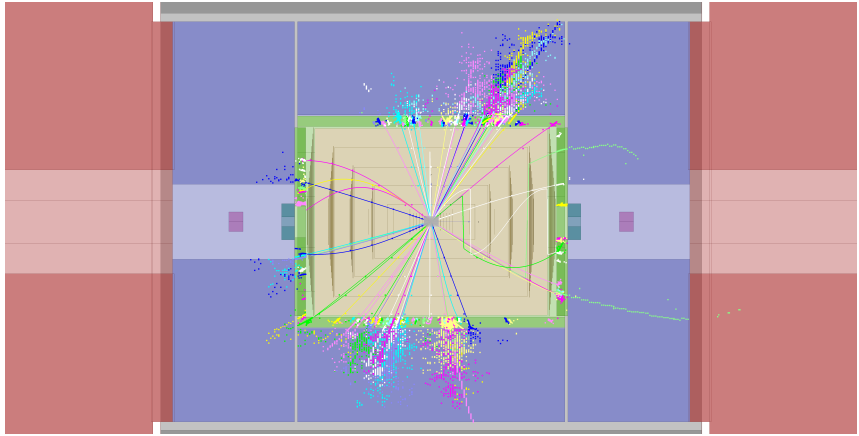}
      \caption{Event display of a t$\bar{\text{t}}$H decay event at $\sqrt{s} = 1.4$\,TeV in the CLIC\_SiD detector. The high granularity of the calorimeter enables tracking of the shower development and can be used as input to a particle flow algorithm. Figure taken from Ref. \cite{clic2016-01}.}
      \label{fig:granularity}
\end{figure}

To resolve the individual shower components and minimise the confusion term, a high cell granularity and precise time information are required (see Fig. \ref{fig:granularity}). The granularity choices for the CLIC detector are $5 \times 5 \Umm^2$ in the electromagnetic calorimeter and $30 \times 30 \Umm^2$ in the hadronic calorimeter. The timing requirements are set to $\sigma_t \approx 1$\,ns at the cell level. Current technological choices are silicon pad sensors and tungsten absorbers for the electromagnetic calorimeter and scintillating tiles with silicon photomultiplier readout and steel absorbers for the hadronic calorimeter. Developments in R\&D and prototyping in this area are pursued within the CALICE collaboration \cite{sefkow2016} and have strong synergies with other projects, such as the planned upgrade of the CMS forward calorimeters \cite{hgc2015}.

\subsubsection{Forward calorimeters}
\begin{figure}
      \centering
      \includegraphics[width=0.95\textwidth]{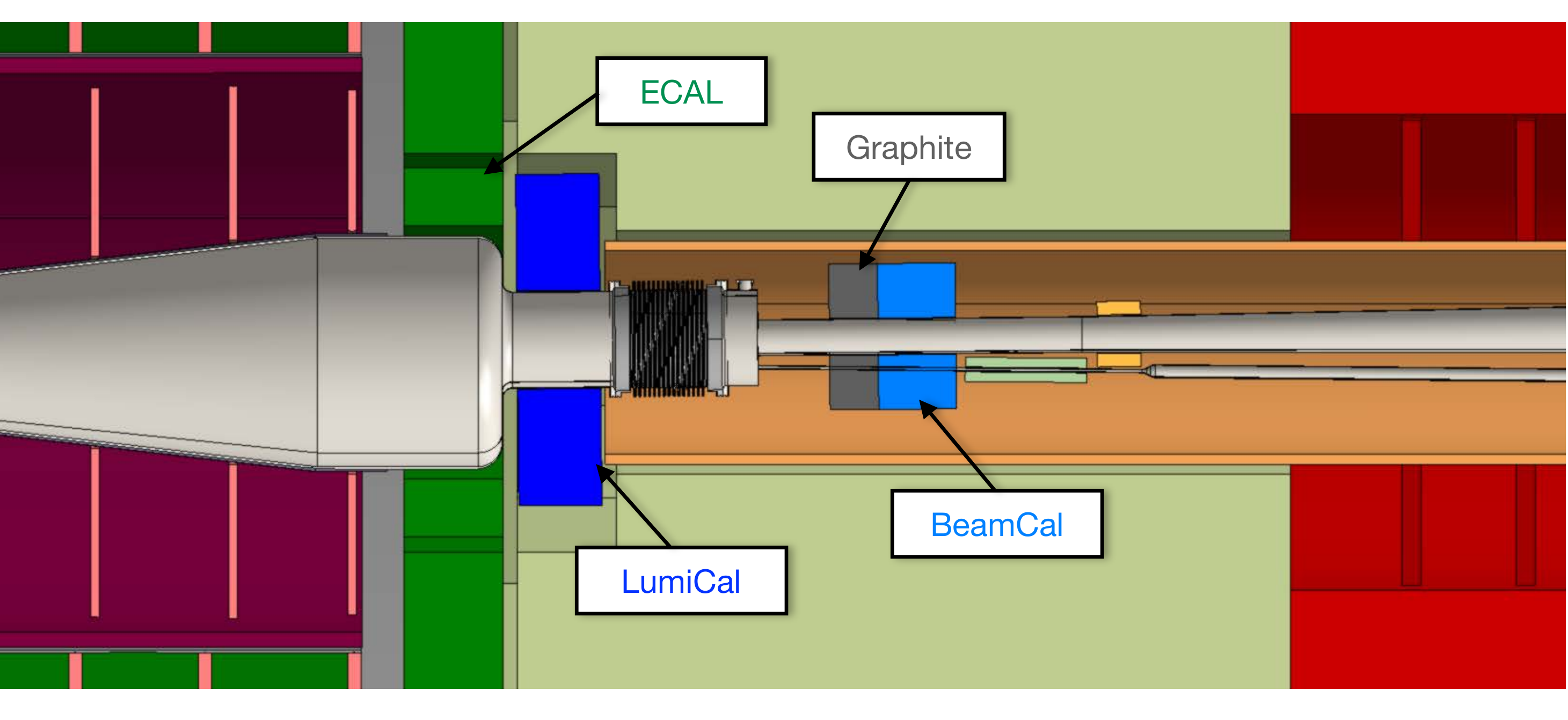}
      \caption{Forward region of the CLIC detector. The BeamCal is used to tag forward-boosted high-energy electrons; the LumiCal delivers a precise luminosity measurement. A block of graphite is used to reduce the flux of back\-scattered particles into the interaction region.}
      \label{fig:forward}
\end{figure}

The very forward region consists of two additional calorimeters that extend to very low angles. Fig. \ref{fig:forward} displays this region. The BeamCal is used for forward tagging of high-energy electrons and can deliver fast luminosity estimation. For precise luminosity measurement, the LumiCal is used. It measures the absolute luminosity via the number of Bhabha scattering events at low angles. To determine the shape of the luminosity spectrum, information on large-angle scattering from the tracker and calorimeter are also used \cite{sailer2013}. The LumiCal and BeamCal cover polar angles of 38--110\,mrad and 10--40\,mrad, respect\-ively. The final focusing magnet QD0 is situated outside the detector. Design efforts for both forward calorimeters are performed within the FCAL collaboration \cite{fcalweb}.

One issue is the flux of backscattered particles from the forward region into the interaction region. To minimise this flux, a graphite block is situated upstream of the BeamCal. For the BeamCal, radiation-hard sensors are also important, as this device will see radiation doses of several MGy in the innermost regions. Both GaAs and CVD diamond sensors are under consideration \cite{fcal2015}.

\subsection{Muon identification system}
The most import task of the muon system is to identify muons with high efficiency and purity. The system is arranged in six layers interleaved in the return yoke. It does not improve the momentum resolution any further but the first layer acts as a tail catcher to improve the jet energy resolution. The time resolution needed is $\sigma_t \approx 1$\,ns and cell sizes of $30 \times 30 \Umm^2$ are used to keep the muon tagging efficiency close to 1 and occupancies at manageable levels. Resistive plate chambers or scintillating tiles with silicon photomultiplier readout are considered to fulfil these requirements.


\section{Conclusions}
The physics goals and accelerator design set challenging requirements on a future CLIC detector. This includes an excellent jet energy and track momentum resolution, efficient flavour tagging capabilities, large geometrical coverage, and efficient suppression of the beam-induced background. A detector concept to meet these requirements has been presented. It is optimised for particle flow calorimetry and builds on fast timing capabilities and high cell granularities throughout the detector. An extensive R\&D programme to show the technical feasibility of the detector design is in progress.


\end{document}